\documentclass[ ,showpacs
 ,secnumarabic, nobibnotes, nofootinbib, aps, prd]{revtex4}
\usepackage{bm}
\usepackage{graphicx}

\begin{document}

\title{Proton structure functions and quark orbital motion\footnote{%
Prepared for the 17th International Spin Physics Symposium, SPIN2006, Kyoto,
Japan, Oct. 2.-7., 2006.}}
\author{Petr Z\'{a}vada}
\email{zavada@fzu.cz}
\affiliation{Institute of Physics, Academy of Sciences of the Czech Republic, \\
Na Slovance 2, CZ-182 21 Prague 8}

\begin{abstract}
Covariant version of the quark-parton model is studied. Dependence of the
structure functions on the 3D quark intrinsic motion is discussed. The
important role of the quark orbital momentum, which is a particular case of
intrinsic motion, appears as a direct consequence of the covariant
description. Effect of orbital motion is substantial especially for
polarized structure functions. At the same time, the procedure for obtaining
the quark momentum distribution from the structure functions is suggested.
\end{abstract}

\maketitle

%\date{August, 2006}

\section{Introduction}

The nucleon structure functions are basic tool for understanding the nucleon
internal structure in the language of QCD. And at the same time, the
measuring and analysis of the structure functions represent the important
experimental test of this theory. Unpolarized nucleon structure functions
are known with high accuracy in very broad kinematical region, but in recent
years also some precision measurements on the polarized structure functions
have been completed \cite{e142,e154,her1,her2,adeva,e143,E155}. For present
status of the nucleon spin structure see e.g. \cite{spin04} and citations
therein. The more formal aspects of the nucleon structure functions are
explained in \cite{efrem}. In fact only the complete set of the four
electromagnetic unpolarized and polarized structure functions $%
F_{1},F_{2},g_{1}$ and $g_{2}$ can give a consistent picture of the nucleon.
However, this picture is usually drawn in terms of the distribution
functions, which are connected with the structure functions by some
model-dependent way. Distribution functions are not directly accessible from
the experiment and model, which is normally applied for their extraction
from the structure functions is the well known quark-parton model (QPM).
Application of this model for analysis and interpretation of the unpolarized
data does not create any contradiction. On the other hand, the situation is
much less  clear in the case of spin functions $g_{1}$ and $g_{2}$.

In our previous study \cite{zav1,zav2} we have suggested, that a 
reasonable explanation of the experimentally measured spin functions $g_{1},$
$g_{2}$ is possible in terms of a generalized covariant QPM, in which the
quark intrinsic motion (i.e. 3D motion with respect to the nucleon rest
frame) is consistently taken into account. Therefore the quark transversal
momentum appears in this approach on the same level as the longitudinal one.
The quarks are represented by the free Dirac spinors, which allows to obtain
exact and covariant solution for relations between the quark momentum
distribution functions and the structure functions accessible from
experiment. In this way the model (in its present leading order version)
contains no dynamics but only \textquotedblleft exact\textquotedblright\
kinematics of quarks, so it can be effective tool for analysis and
interpretation of the experimental data on structure functions, particularly
for separating effects of the dynamics (QCD) from effects of the kinematics.
This point of view is well supported by our previous results:

\textit{a)} In the cited papers we showed, that the model simply implies the
well known sum rules (Wanzura-Wilczek, Efremov-Leader-Teryaev,
Burkhardt-Cottingham) for the spin functions $g_{1},$ $g_{2}$.

\textit{b}) Simultaneously, we showed that the same set of assumptions
implies rather substantial dependence of the first moment $\Gamma _{1}$ of
the function $g_{1}$ on the kinematical effects.

\textit{c)} Further, we showed that the model allows to calculate the
functions $g_{1},$ $g_{2}$ from the unpolarized valence quark distributions
and the result is quite compatible with the experimental data.

\textit{d)} In the paper \cite{tra}\ we showed that the model allows to
relate the transversity distribution to some other structure functions.

These results cannot be obtained from the standard versions of the QPM
(naive or the QCD improved), which are currently used for the analysis of
experimental data on structure functions. The reason is, that the standard
QPM is based on the simplified and non-covariant kinematics in the infinite
momentum frame (IMF), which does not allow to properly take into account the
quark intrinsic or orbital motion.

The subject of our previous study was the question: \textit{What is the
dependence of the structure functions on quark intrinsic motion?} The aim of
the present paper is a discussion of related problems:

\textit{1. How to extract information about the quark intrinsic motion from
the experimentally measured structure functions?}

\textit{2. What is the role of the quark orbital momentum, which is a
particular case of intrinsic motion?}

The paper is organized as follows. In the first part of Sec. \ref{sec2} the
basic formulas, which follow from the generalized QPM, are presented.
Resulting general\ covariant relations are compared with their limiting
case, which is represented by the standard formulation of the QPM in the
IMF. In the next part of the section the relations for calculation of 3D
quark momentum distributions from the structure functions are derived. The
quark momentum distributions are obtained from the experimentally measured
structure functions $F_{2}$ and $g_{1}$ \ and it is shown how their
combination allows to calculate the momentum distributions of the positively
and negatively polarized quarks. The particular form of the quark intrinsic
motion is the orbital momentum. In Sec. \ref{sec3} the role of the quark
orbital momentum in covariant description is discussed and it is shown, why
its contribution to the total quark angular momentum can be quite
substantial. It is demonstrated, that the orbital motion is an inseparable
part of the covariant approach. The problem of quark orbital momentum in the
context of nucleon spin was recognized and studied also in many previous
papers, see e.g. \cite%
{sehgal,ratc,abbas,kep,casu,bqma,brod,waka,waka1,song,ji}. The last section
is devoted to a short summary and conclusion.

\section{Structure functions and intrinsic quark motion}

\label{sec2}In our previous study \cite{zav0,zav1,zav2} of the proton
structure functions we showed, how these functions depend on the intrinsic
motion of quarks. The quarks in the suggested model are represented by the
free fermions, which are in the proton rest frame described by the set of
distribution functions with spheric symmetry $G_{k}^{\pm }(p_{0})d^{3}p$,
where $p_{0}=\sqrt{m^{2}+\mathbf{p}^{2}}$ and symbol $k$ represents the
quark and antiquark flavors. These distributions measure the probability to
find a quark of given flavor in the state%
\begin{equation}
u\left( \mathbf{p},\lambda \mathbf{n}\right) =\frac{1}{\sqrt{N}}\left( 
\begin{array}{c}
\phi _{\lambda \mathbf{n}} \\ 
\frac{\mathbf{p}\mathbf{\sigma }}{p_{0}+m}\phi _{\lambda \mathbf{n}}%
\end{array}%
\right) ;\qquad \frac{1}{2}\mathbf{n\sigma }\phi _{\lambda \mathbf{n}%
}=\lambda \phi _{\lambda \mathbf{n}},\qquad N=\frac{2p_{0}}{p_{0}+m},
\label{sp8}
\end{equation}%
where $m$ and $p$\ are the quark mass and momentum, $\lambda =\pm 1/2,$ $%
\phi _{\lambda \mathbf{n}}^{\dagger }\phi _{\lambda \mathbf{n}}=1$ and$\ 
\mathbf{n}$ coincides with the direction of proton polarization. The
distributions with the corresponding quark (and antiquark) charges $e_{k}$
allow to define the generic functions $G$ and $\Delta G$\footnote{%
In the papers \cite{zav1,zav2} we used different notation for the
distributions defined by Eqs.(\ref{sp8a}) and (\ref{sp9}): $G_{k}^{\pm
},\Delta G_{k}$ and $\Delta G$ were denoted as $h_{k\pm },\Delta h_{k}$ and $%
H$. Apart of that we assumed for simplicity that only three (valence) quarks
contribute to the sums (\ref{sp8a}) and (\ref{sp9}). In present paper we
assume contribution of all the quarks and antiquarks, but apparently general
form of the relations like (\ref{cra31}) - (\ref{cr31}) is independent of
chosen set of quarks.},%
\begin{equation}
G(p_{0})=\sum_{k}e_{k}^{2}G_{k}(p_{0}),\quad G_{k}(p_{0})\equiv
G_{k}^{+}(p_{0})+G_{k}^{-}(p_{0}),  \label{sp8a}
\end{equation}%
\begin{equation}
\Delta G(p_{0})=\sum_{k}e_{k}^{2}\Delta G_{k}(p_{0}),\quad \Delta
G_{k}(p_{0})\equiv G_{k}^{+}(p_{0})-G_{k}^{-}(p_{0}),  \label{sp9}
\end{equation}%
from which the structure functions can be obtained. If $q$ is momentum of
the photon absorbed by the proton of the momentum $P$ and mass $M,$ in which
the phase space of quarks is controlled by the distributions $G_{k}^{\pm
}(p_{0})d^{3}p$, then there are the following representations of
corresponding structure functions.

\textit{A. Manifestly covariant representation}

\textit{i) unpolarized structure functions:} 
\begin{equation}
F_{1}(x)=\frac{M}{2}\left( A+\frac{B}{\gamma }\right) ,\qquad F_{2}(x)=\frac{%
Pq}{2M\gamma }\left( A+\frac{3B}{\gamma }\right) ,  \label{sp9b}
\end{equation}%
where%
\begin{equation}
A=\frac{1}{Pq}\int G\left( \frac{Pp}{M}\right) \left[ pq-m^{2}\right] \delta
\left( \frac{pq}{Pq}-x\right) \frac{d^{3}p}{p_{0}},  \label{sp9c}
\end{equation}%
\begin{equation}
B=\frac{1}{Pq}\int G\left( \frac{pP}{M}\right) \left[ \left( \frac{Pp}{M}%
\right) ^{2}+\frac{\left( Pp\right) \left( Pq\right) }{M^{2}}-\frac{pq}{2}%
\right] \delta \left( \frac{pq}{Pq}-x\right) \frac{d^{3}p}{p_{0}}
\label{sp9d}
\end{equation}%
and%
\begin{equation}
\gamma =1-\left( \frac{Pq}{Mq}\right) ^{2}.  \label{sp9e}
\end{equation}%
The functions $F_{1}=MW_{1}$ and $F_{2}=\left( Pq/M\right) W_{2}$ follow
from the tensor equation%
\begin{eqnarray}
&&\left( -g_{\alpha \beta }+\frac{q_{\alpha }q_{\beta }}{q^{2}}\right)
W_{1}+\left( P_{\alpha }-\frac{Pq}{q^{2}}q_{\alpha }\right) \left( P_{\beta
}-\frac{Pq}{q^{2}}q_{\beta }\right) \frac{W_{2}}{M^{2}}  \label{sp9f} \\
&=&\int G\left( \frac{pP}{M}\right) \left[ 2p_{\alpha }p_{\beta }+p_{\alpha
}q_{\beta }+q_{\alpha }p_{\beta }-g_{\alpha \beta }pq\right] \delta \left(
\left( p+q\right) ^{2}-m^{2}\right) \frac{d^{3}p}{p_{0}}.  \nonumber
\end{eqnarray}%
After modification of the delta function term%
\begin{equation}
\delta \left( \left( p+q\right) ^{2}-m^{2}\right) =\delta \left(
2pq+q^{2}\right) =\delta \left( 2Pq\left( \frac{pq}{Pq}-\frac{Q^{2}}{2Pq}%
\right) \right) =\frac{1}{2Pq}\delta \left( \frac{pq}{Pq}-x\right) ;\qquad
q^{2}=-Q^{2},\quad x=\frac{Q^{2}}{2Pq},  \label{sp9g}
\end{equation}%
the dependence on the Bjorken $x$ is introduced. Then contracting with the
tensors $g_{\alpha \beta }$ and $P_{\alpha }P_{\beta }$ gives the set of two
equations, which determine the functions $F_{1},F_{2}$ in accordance with
Eqs. (\ref{sp9b})-(\ref{sp9e}).

\textit{ii) polarized structure functions:}

As follows from \cite{zav1} the corresponding spin functions in covariant
form read%
\begin{equation}
g_{1}=Pq\left( G_{S}-\frac{Pq}{qS}G_{P}\right) ,\qquad g_{2}=\frac{\left(
Pq\right) ^{2}}{qS}G_{P},  \label{cra31}
\end{equation}%
where $S$ is the proton spin polarization vector and the functions $%
G_{P},G_{S}$\ are defined as%
\begin{equation}
G_{P}=\frac{m}{2Pq}\int \Delta G\left( \frac{pP}{M}\right) \frac{pS}{pP+mM}%
\left[ 1+\frac{1}{mM}\left( pP-\frac{pu}{qu}Pq\right) \right] \delta \left( 
\frac{pq}{Pq}-x\right) \frac{d^{3}p}{p_{0}},  \label{cr30}
\end{equation}%
\begin{equation}
G_{S}=\frac{m}{2Pq}\int \Delta G\left( \frac{pP}{M}\right) \left[ 1+\frac{pS%
}{pP+mM}\frac{M}{m}\left( pS-\frac{pu}{qu}qS\right) \right] \delta \left( 
\frac{pq}{Pq}-x\right) \frac{d^{3}p}{p_{0}};  \label{cr31}
\end{equation}%
\[
u=q+\left( qS\right) S-\frac{\left( Pq\right) }{M^{2}}P.
\]

\textit{B. Rest frame representation for} $Q^{2}\gg 4M^{2}x^{2}$

As follows from the Appendix in \cite{zav1}, if $Q^{2}\gg 4M^{2}x^{2}$ and
the above integrals are expressed in terms of the proton rest frame
variables, then one can substitute 
\[
\frac{pq}{Pq}\rightarrow \frac{p_{0}+p_{1}}{M}
\]%
and the structure functions are simplified as:%
\begin{eqnarray}
F_{1}(x) &=&\frac{Mx}{2}\int G(p_{0})\delta \left( \frac{p_{0}+p_{1}}{M}%
-x\right) \frac{d^{3}p}{p_{0}},  \label{sp9h} \\
F_{2}(x) &=&Mx^{2}\int G(p_{0})\delta \left( \frac{p_{0}+p_{1}}{M}-x\right) 
\frac{d^{3}p}{p_{0}},  \label{sp9a} \\
g_{1}(x) &=&\frac{1}{2}\int \Delta G(p_{0})\left( m+p_{1}+\frac{p_{1}^{2}}{%
p_{0}+m}\right) \delta \left( \frac{p_{0}+p_{1}}{M}-x\right) \frac{d^{3}p}{%
p_{0}},  \label{sp10} \\
g_{2}(x) &=&-\frac{1}{2}\int \Delta G(p_{0})\left( p_{1}+\frac{%
p_{1}^{2}-p_{T}^{2}/2}{p_{0}+m}\right) \delta \left( \frac{p_{0}+p_{1}}{M}%
-x\right) \frac{d^{3}p}{p_{0}},  \label{sp11}
\end{eqnarray}%
where the $p_{1}$ and $p_{T}$ are longitudinal and transversal quark
momentum components.

\textit{C. Standard IMF representation}

The usual formulation of the QPM gives the known relations between the
structure and distribution functions \cite{efrem}:%
\begin{equation}
F_{1}(x)=\frac{1}{2}\sum_{q}e_{q}^{2}q(x),\qquad
F_{2}(x)=x\sum_{q}e_{q}^{2}q(x),  \label{sp11a}
\end{equation}%
\begin{equation}
g_{1}(x)=\frac{1}{2}\sum_{q}e_{q}^{2}\Delta q(x),\qquad g_{2}(x)=0,
\label{sp11b}
\end{equation}%
where%
\begin{equation}
q(x)=q^{+}(x)+q^{-}(x),\qquad \Delta q(x)=q^{+}(x)-q^{-}(x).  \label{sp11c}
\end{equation}
In the Appendix \ref{app0} we have proved that these relations represent the
particular, limiting case of the covariant relations (\ref{sp9b}) and (\ref%
{cra31}).

The three versions of the relations between the structure functions and the
quark distributions can be compared:

\textit{a)} If we skip the function $g_{2}$ in the version \textit{C}, then
the relations (\ref{sp11a}) and (\ref{sp11b}) practically represent identity
between the structure functions and quark distributions. Such simple
relations are valid only for the IMF approach based on the approximation (%
\ref{b2}), which means that the quark intrinsic motion is suppressed. In
more general versions \textit{A} and \textit{B}, where the intrinsic motion
is allowed, the relations are more complex. The intrinsic motion strongly
modifies also the $g_{2}$. In the version \textit{C} there is $g_{2}(x)=0$,
but $g_{2}(x)\neq 0$ in the \textit{A} and \textit{B}.

\textit{b)} The version \textit{B} allows to easily calculate the
(substantial) dependence of the first moment $\Gamma _{1}$ on the rate of
intrinsic motion. A more detailed discussion follows in the next section.
The same approach implies that functions $g_{1}$ and $g_{2}$ for massless
quarks satisfy the relation equivalent to the Wanzura-Wilczek term and obey
some well known sum rules, that is shown in \cite{zav1}.

\textit{c)} The functions $F_{1}$ and $F_{2}$ exactly satisfy the
Callan-Gross relation $F_{2}(x)/F_{1}(x)=2x$ in the versions \textit{B} and
C, but this relation is satisfied only approximately in the \textit{A}: $%
F_{2}(x)/F_{1}(x)\approx 2x+O\left( 4M^{2}x^{2}/Q^{2}\right) $.

The task which was solved in different approximations above can be
formulated: How to obtain the structure functions $F_{1},F_{2}$ and $%
g_{1},g_{2}$ from the probabilistic distributions $G$ and $\Delta G$ defined
by Eqs. (\ref{sp8a}) and (\ref{sp9})? In the next we will study the inverse
task, the aim is to find out a rule for obtaining the distribution functions 
$G$ and $\Delta G$ from the structure functions. In the present paper we
consider the functions $F_{2}$ and $g_{1}$ represented by Eqs. (\ref{sp9a})
and (\ref{sp10}). As follows from the Appendix A in \cite{zav2}, the
function%
\begin{equation}
V_{n}(x)=\int K(p_{0})\left( \frac{p_{0}}{M}\right) ^{n}\delta \left( \frac{%
p_{0}+p_{1}}{M}-x\right) d^{3}p  \label{gp1}
\end{equation}%
satisfies%
\begin{equation}
V_{n}^{\prime }(x_{\pm })x_{\pm }=\mp 2\pi MK(\xi )\xi \sqrt{\xi ^{2}-m^{2}}%
\left( \frac{\xi }{M}\right) ^{n};\qquad x_{\pm }=\frac{\xi \pm \sqrt{\xi
^{2}-m^{2}}}{M}.  \label{gp2}
\end{equation}%
In this section we consider only the case $m\rightarrow 0$, then%
\begin{equation}
V_{n}^{\prime }(x)x=-2\pi MK(\xi )\xi ^{2}\left( \frac{\xi }{M}\right)
^{n};\qquad x=\frac{2\xi }{M}.  \label{gp3}
\end{equation}%
As we shall see below, with the use of this relation one can obtain the
probabilistic distributions $G$ and $\Delta G$ from the experimentally
measured structure functions.

Let us remark that in present stage the QCD evolution is not included into
the model. However, this fact does not represent any restriction for the
present purpose - to obtain information about distributions of quarks at
some $Q^{2}$ from the structure functions measured at the same $Q^{2}$.
Distribution of the gluons is another part of the proton picture. But since
our present discussion is directed to the relation between the structure
functions and corresponding (quark) distributions at given scale, the gluon
distribution is left aside.

\subsection{Momentum distribution from structure function $F_{2}$}

In an accordance with the definition (\ref{gp1}) in which the distribution $%
K(p_{0})$\ is substituted\ by the $G(p_{0})$, the structure function (\ref%
{sp9a}) can be written in the form

\begin{equation}
F_{2}(x)=x^{2}V_{-1}(x).  \label{gp5}
\end{equation}%
Then, with the use of the relation (\ref{gp3}) one gets%
\begin{equation}
G(p)=-\frac{1}{\pi M^{3}}\left( \frac{F_{2}(x)}{x^{2}}\right) ^{\prime }=%
\frac{1}{\pi M^{3}x^{2}}\left( \frac{2F_{2}(x)}{x}-F_{2}^{\prime }(x)\right)
;\qquad x=\frac{2p}{M},\qquad p\equiv \sqrt{\mathbf{p}^{2}}=p_{0}.
\label{gp8}
\end{equation}%
Probability distribution $G$ measures number of quarks in the element $%
d^{3}p $. Since $d^{3}p=4\pi p^{2}dp$, the distribution measuring the number
of quarks in the element $dp/M$ reads%
\begin{equation}
4\pi p^{2}MG(p)=-x^{2}\left( \frac{F_{2}(x)}{x^{2}}\right) ^{\prime }=\frac{%
2F_{2}(x)}{x}-F_{2}^{\prime }(x);\qquad x=\frac{2p}{M}.  \label{gp9}
\end{equation}%
Let us note, the maximum value of quark momentum is $p_{\max }=M/2$, which
is a consequence of the kinematics in the proton rest frame, where the
single quark momentum must be compensated by the momentum of the other
partons.

\begin{figure}[tbp]
\includegraphics[width=18cm]{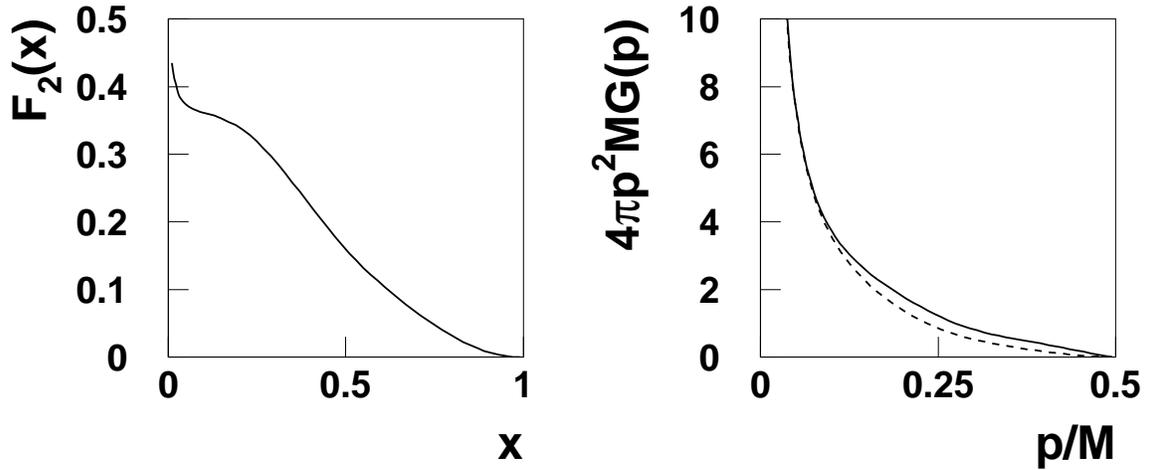}
\caption{Proton structure function $F_{2}$ at $Q^{2}=4GeV^{2}$ (left).
Corresponding calculation of the quark momentum distributions in the proton
rest frame: the $p$ and $p_{T}$ distributions are represented \ by solid and
dashed lines (right).}
\label{fgr1}
\end{figure}

Another quantity, which can be obtained, is the distribution of the quark
transversal momentum. Obviously the integral%
\begin{equation}
\frac{dN}{dp_{T}^{2}}=\int G(p)\delta \left(
p_{2}^{2}+p_{3}^{2}-p_{T}^{2}\right) d^{3}p,  \label{gp12}
\end{equation}%
which measures the number of quarks in the element $dp_{T}^{2}$, can be
modified as%
\begin{equation}
\frac{dN}{dp_{T}^{2}}=2\pi \int_{0}^{\sqrt{p_{\max }^{2}-p_{T}^{2}}}G\left( 
\sqrt{p_{1}^{2}+p_{T}^{2}}\right) dp_{1}.  \label{gp13}
\end{equation}%
It follows, that the distribution measuring number of quarks in the element $%
dp_{T}/M$ reads:%
\begin{equation}
P(p_{T})=M\frac{dN}{dp_{T}}=4\pi p_{T}M\int_{0}^{\sqrt{p_{\max
}^{2}-p_{T}^{2}}}G\left( \sqrt{p_{1}^{2}+p_{T}^{2}}\right) dp_{1}.
\label{gp14}
\end{equation}%
Let us point out that the distribution $P(p_{T}),$ equally as the
distributions $G$ and $\Delta G,$ represents the combination of the
distributions related to different quark and antiquark flavors - like in
Eqs. (\ref{sp8a}) or (\ref{sp9}). With the use of Eq. (\ref{gp8}) one gets
distributions%
\begin{equation}
P(p_{T})=\frac{4p_{T}}{M^{2}}\int_{0}^{\sqrt{p_{\max }^{2}-p_{T}^{2}}}\frac{1%
}{x^{2}}\left( \frac{2F_{2}(x)}{x}-F_{2}^{\prime }(x)\right) dp_{1};\qquad x=%
\frac{2\sqrt{p_{1}^{2}+p_{T}^{2}}}{M}.  \label{gp16}
\end{equation}%
In Fig. \ref{fgr1} \ the structure function $F_{2}$ together with the
corresponding distributions calculated with the use of relations (\ref{gp9})
and (\ref{gp16}) are displayed. 
\ For the proton structure function the phenomenological fit performed in
the Appendix of the paper \cite{adeva} was used. Using the  Eq. (\ref{gp8})
one can calculate the mean value%
\begin{equation}
\left\langle p\right\rangle =\frac{\int pG(p)d^{3}p}{\int G(p)d^{3}p}=\frac{M%
}{2}\frac{\int_{0}^{1}x^{3}\left( \frac{F_{2}(x)}{x^{2}}\right) ^{\prime }dx%
}{\int_{0}^{1}x^{2}\left( \frac{F_{2}(x)}{x^{2}}\right) ^{\prime }dx},
\label{gp16a}
\end{equation}%
but since the extrapolation of the structure function for $x\rightarrow 0$
gives in the denominator divergent integral, it follows that $\left\langle
p\right\rangle \rightarrow 0$. Nontrivial value can be obtained with the
integration cutoff $x>x_{\min }$.

\subsection{Momentum distribution from structure function $g_{1}$}

\begin{figure}[tbp]
\includegraphics[width=18cm]{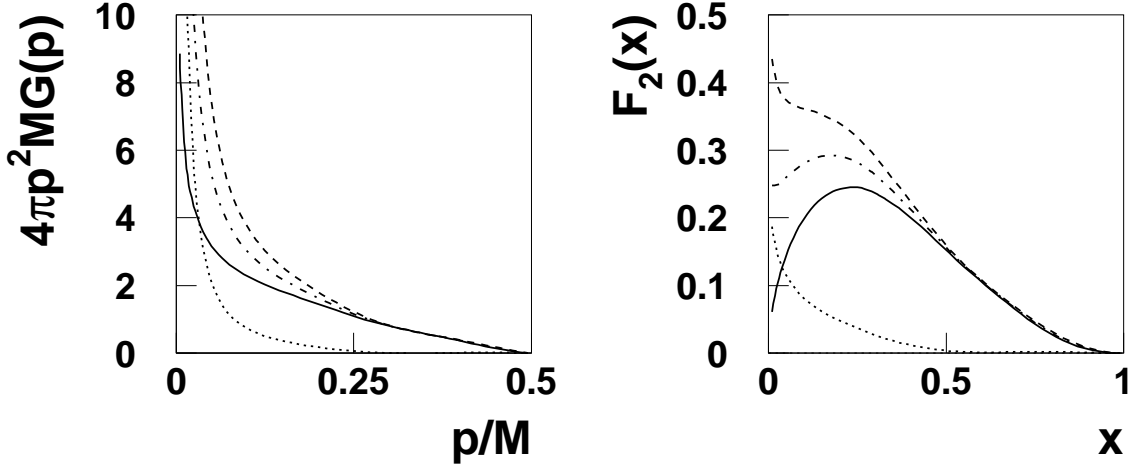}
\caption{Probability distributions $\Delta G,G,G^{+}$ and $G^{-}$ are
represented by solid, dashed, dash-and-dot and dotted lines (left) and
corresponding structure functions $\Delta F_{2},F_{2},F_{2}^{+}$ and $%
F_{2}^{-}$ \ (right).}
\label{fgr2}
\end{figure}

Now, we shall determine the distribution $\Delta G$ defined in Eq. (\ref{sp9}%
) from the spin function $g_{1}$\ - similarly as distribution $G$ was
obtained from the function $F_{2}$. In the paper \cite{zav2}, Eq. (44), we
proved that%
\begin{equation}
g_{1}(x)=V_{0}(x)-\int_{x}^{1}\left( 4\frac{x^{2}}{y^{3}}-\frac{x}{y^{2}}%
\right) V_{0}(y)dy,  \label{gp17}
\end{equation}%
where the function $V_{0}$ is defined by Eq. (\ref{gp1}) for $n=0$ and $%
K(p)= $ $\Delta G(p)$. In the Appendix \ref{app1} it is shown, that the last
relation can be modified to:%
\begin{equation}
V_{-1}(x)=\frac{2}{x}\left( g_{1}(x)+2\int_{x}^{1}\frac{g_{1}(y)}{y}%
)dy\right) .  \label{GP19}
\end{equation}%
Then, in an accordance with Eq. (\ref{gp3}), we obtain 
\begin{equation}
V_{-1}^{\prime }(x)=-\pi M^{3}\Delta G(p);\qquad x=\frac{2p}{M},
\label{gp20}
\end{equation}%
so the last two relations imply 
\begin{equation}
\Delta G(p)=\frac{2}{\pi M^{3}x^{2}}\left( 3g_{1}(x)+2\int_{x}^{1}\frac{%
g_{1}(y)}{y})dy-xg_{1}^{\prime }(x)\right) ;\qquad x=\frac{2p}{M}.
\label{gp21}
\end{equation}%
Obviously this distribution together with the distribution (\ref{gp8})
allows to obtain the polarized distributions $G^{\pm }$ as%
\begin{equation}
G^{\pm }(p)=\sum_{q}e_{q}^{2}G_{q}^{\pm }(p)=\frac{1}{2}\left( G(p)\pm
\Delta G(p)\right) .  \label{gp22}
\end{equation}%
These distributions can be with the use of Eqs. (\ref{gp8}) and (\ref{gp21})
obtained from experimental data on the $F_{2}$ and \ $g_{1}$.The result is
displayed in the left part of Fig. \ref{fgr2}, for which the function $g_{1}$
was parameterized by the fit of the world data \cite{E155} at $%
Q^{2}=4GeV^{2} $. 

With the use of \ the relation (\ref{sp9a}) one can formally calculate the
partial structure functions corresponding to the subsets of positively and
negatively polarized quarks:%
\begin{equation}
F_{2}^{\pm }(x)=Mx^{2}\int G^{\pm }(p)\delta \left( \frac{p_{0}+p_{1}}{M}%
-x\right) \frac{d^{3}p}{p_{0}}.  \label{gp25bAD}
\end{equation}%
Apparently it holds%
\begin{equation}
F_{2}(x)=F_{2}^{+}(x)+F_{2}^{-}(x)  \label{gad1}
\end{equation}%
and one can define also%
\begin{equation}
\Delta F_{2}(x)=F_{2}^{+}(x)-F_{2}^{-}(x)  \label{gad2}
\end{equation}%
or equivalently%
\begin{equation}
\Delta F_{2}(x)=Mx^{2}\int \Delta G(p)\delta \left( \frac{p_{0}+p_{1}}{M}%
-x\right) \frac{d^{3}p}{p_{0}}.  \label{gp23}
\end{equation}%
This equality can be written as%
\begin{equation}
\Delta F_{2}(x)=x^{2}V_{-1}(x),  \label{gp24}
\end{equation}%
which after inserting from Eq. (\ref{GP19}) gives%
\begin{equation}
\Delta F_{2}(x)=2x\left( g_{1}(x)+2\int_{x}^{1}\frac{g_{1}(y)}{y})dy\right) .
\label{gp25}
\end{equation}%
The structure functions $F_{2},\Delta F_{2},F_{2}^{\pm }$ are shown in the
right part of Fig. \ref{fgr2} and one can observe:

\textit{i)} Shape of the function $\Delta F_{2}$ implies, that contributions
of oppositely polarized quarks in Eq. (\ref{gad2}) are canceled out in the
region of low $x$. In fact the shape is similar to that of the function $%
F_{2val}$ corresponding to the valence quarks. This suggests that a dominant
spin contribution comes from the valence region.

\textit{ii)} The distributions $\Delta G,G$ and $G^{+}$ are very close
together in the region of higher momenta and simultaneously $G^{-}$ is close
to zero in the same region. And the same holds for corresponding structure
functions $\Delta F_{2},F_{2}$ and $F_{2}^{\pm }$ in the higher $x$ region.
In an accordance with the definitions (\ref{sp8a}),(\ref{sp9}) it follows,
that%
\begin{equation}
G_{q}^{+}(p)\simeq G_{q}(p),\qquad G_{q}^{-}(p)\simeq 0.  \label{gp29}
\end{equation}%
In the other words, polarization of quarks or partons with higher intrinsic
energy (and/or higher $x$) coincides with the proton polarization.

The mean value of the distribution $\Delta G$ can be estimated as%
\begin{equation}
\left\langle p\right\rangle =\frac{\int p\Delta G(p)d^{3}p}{\int \Delta
G(p)d^{3}p}=\frac{M}{2}\left\langle x\right\rangle ;\qquad \left\langle
x\right\rangle =\frac{\int_{0}^{1}xg_{1}(x)dx}{\int_{0}^{1}g_{1}(x)dx}.
\label{GP26}
\end{equation}%
The proof of this relation is done in the Appendix \ref{app2}. The numerical
calculation with the $g_{1}$\ fit gives $\left\langle p\right\rangle
=0.113GeV/c$. However, interpretation of this average momentum should be
done with some care since $\Delta G_{q}$ can be negative, in principle.
Average momentum $\left\langle p\right\rangle $ allows to calculate the mean
transversal momentum; $\left\langle p_{T}\right\rangle =\pi /4\cdot
\left\langle p\right\rangle $.

\section{Intrinsic quark motion and orbital momentum}

\label{sec3}The rule of quantum mechanics says, that angular momentum
consists of the orbital and spin part \ $\mathbf{j=l+s}$\ and that in the
relativistic case the $\mathbf{l}$ and $\mathbf{s}$\ are not conserved
separately, but only the total angular momentum $j$\ is conserved.\textit{\ }%
This simple fact was in the context of quarks inside the nucleon pointed out
in \cite{liang}. It means, that only $j^{2},j_{z}$ are well-defined quantum
numbers and corresponding states of the particle with spin $1/2$ are
represented by the bispinor spherical waves \cite{lali}%
\begin{equation}
\psi _{kjlj_{z}}\left( \mathbf{p}\right) =\frac{\delta (p-k)}{p\sqrt{2p_{0}}}%
\left( 
\begin{array}{c}
i^{-l}\sqrt{p_{0}+m}\Omega _{jlj_{z}}\left( \mathbf{\omega }\right) \\ 
i^{-\lambda }\sqrt{p_{0}-m}\Omega _{j\lambda j_{z}}\left( \mathbf{\omega }%
\right)%
\end{array}%
\right) ,  \label{sp19}
\end{equation}%
where $\mathbf{\omega }=\mathbf{p}/p,$\ $l=j\pm \frac{1}{2},\ \lambda =2j-l$
($l$ defines the parity) and 
\begin{eqnarray*}
\Omega _{j,l,j_{z}}\left( \mathbf{\omega }\right) &=&\left( 
\begin{array}{c}
\sqrt{\frac{j+j_{z}}{2j}}Y_{l,j_{z}-1/2}\left( \mathbf{\omega }\right) \\ 
\sqrt{\frac{j-j_{z}}{2j}}Y_{l,j_{z}+1/2}\left( \mathbf{\omega }\right)%
\end{array}%
\right) ;\quad l=j-\frac{1}{2}, \\
\Omega _{j,l,j_{z}}\left( \mathbf{\omega }\right) &=&\left( 
\begin{array}{c}
-\sqrt{\frac{j-j_{z}+1}{2j+2}}Y_{l,j_{z}-1/2}\left( \mathbf{\omega }\right)
\\ 
\sqrt{\frac{j+j_{z}+1}{2j+2}}Y_{l,j_{z}+1/2}\left( \mathbf{\omega }\right)%
\end{array}%
\right) ;\quad l=j+\frac{1}{2}.
\end{eqnarray*}%
States are normalized as:%
\begin{equation}
\int \psi _{k^{\prime }j^{\prime }l^{\prime }j_{z}^{\prime }}^{\dagger
}\left( \mathbf{p}\right) \psi _{kjlj_{z}}\left( \mathbf{p}\right)
d^{3}p=\delta (k-k^{\prime })\delta _{jj^{\prime }}\delta _{ll^{\prime
}}\delta _{j_{z}j_{z}^{\prime }}.  \label{sp20}
\end{equation}%
The wavefunction (\ref{sp19}) is simplified for $j=j_{z}=1/2$ and $l=0$.
Taking into account that%
\[
Y_{00}=\frac{1}{\sqrt{4\pi }},\qquad Y_{10}=i\sqrt{\frac{3}{4\pi }}\cos
\theta ,\qquad Y_{11}=-i\sqrt{\frac{3}{8\pi }}\sin \theta \exp \left(
i\varphi \right) , 
\]%
one gets:%
\begin{equation}
\psi _{kjlj_{z}}\left( \mathbf{p}\right) =\frac{\delta (p-k)}{p\sqrt{8\pi
p_{0}}}\left( 
\begin{array}{c}
\sqrt{p_{0}+m}\left( 
\begin{array}{c}
1 \\ 
0%
\end{array}%
\right) \\ 
-\sqrt{p_{0}-m}\left( 
\begin{array}{c}
\cos \theta \\ 
\sin \theta \exp \left( i\varphi \right)%
\end{array}%
\right)%
\end{array}%
\right) .  \label{sp22}
\end{equation}%
Let us note, that $j=1/2$ is the minimum angular momentum for the particle
with spin $1/2.$ If one consider the quark state as a superposition%
\begin{equation}
\Psi \left( \mathbf{p}\right) =\int a_{k}\psi _{kjlj_{z}}\left( \mathbf{p}%
\right) dk;\quad \int a_{k}^{\star }a_{k}dk=1  \label{sp21}
\end{equation}%
then its average spin contribution to the total angular momentum reads:%
\begin{equation}
\left\langle s\right\rangle =\int \Psi ^{\dagger }\left( \mathbf{p}\right)
\Sigma _{z}\Psi \left( \mathbf{p}\right) d^{3}p;\qquad \Sigma _{z}=\frac{1}{2%
}\left( 
\begin{array}{cc}
\sigma _{z} & \cdot \\ 
\cdot & \sigma _{z}%
\end{array}%
\right) .  \label{sp26}
\end{equation}%
After inserting from Eqs. (\ref{sp22}), (\ref{sp21}) into (\ref{sp26}) one
gets

\begin{equation}
\left\langle s\right\rangle =\int a_{p}^{\star }a_{p}\frac{\left(
p_{0}+m\right) +\left( p_{0}-m\right) \left( \cos ^{2}\theta -\sin
^{2}\theta \right) }{16\pi p^{2}p_{0}}d^{3}p=\frac{1}{2}\int a_{p}^{\star
}a_{p}\left( \frac{1}{3}+\frac{2m}{3p_{0}}\right) dp.  \label{sp24}
\end{equation}%
Since\ $j=1/2$, the last relation implies for the quark orbital momentum:%
\begin{equation}
\left\langle l\right\rangle =\frac{1}{3}\int a_{p}^{\star }a_{p}\left( 1-%
\frac{m}{p_{0}}\right) dp.  \label{sp25}
\end{equation}%
It means that for quarks in the state $j=j_{z}=1/2$ there are the extreme
scenarios:

\textit{i)} Massive and static quarks ($p_{0}=m$), which implies $%
\left\langle s\right\rangle =j=1/2$ and $\left\langle l\right\rangle =0$.\
This is evident, since without kinetic energy no orbital momentum can be
generated.

\textit{ii)}\ Massless quarks $\left( m\ll p_{0}\right) $, which implies $%
\left\langle s\right\rangle =1/6$ and $\left\langle l\right\rangle =1/3$.

Generally, for $p_{0}\geq m$, one gets $1/3\leq \left\langle s\right\rangle
/j\leq 1.$ In other words, for the states with $p_{0}>m$ part of the total
angular momentum $j=1/2$ is necessarily generated by the orbital momentum.
This is a consequence of quantum mechanics, and not a consequence of the
particular model. For instance, if one assumes the quark effective mass of
the order thousandths and momentum of the order of tenth of $GeV$, then the
second scenario is clearly preferred. Further, the mean kinetic energy
corresponding to the superposition (\ref{sp21}) reads%
\begin{equation}
\left\langle E_{kin}\right\rangle =\int a_{p}^{\star }a_{p}E_{kin}dp;\qquad
E_{kin}=p_{0}-m  \label{gp30}
\end{equation}%
and at the same time the Eq. (\ref{sp25}) can be rewritten as%
\begin{equation}
\left\langle l\right\rangle =\frac{1}{3}\int a_{p}^{\star }a_{p}\frac{E_{kin}%
}{p_{0}}dp.  \label{gp31}
\end{equation}%
It is evident, that for fixed $j=1/2$ both the quantities are in the proton
rest frame almost equivalent: more kinetic energy generates more orbital
momentum and vice versa.

Further, the average spin part $\left\langle s\right\rangle $ of the total
angular momentum $j=1/2$ related to single quark according to Eq. (\ref{sp24}%
) can be compared with the integral%
\begin{equation}
\Gamma _{1}=\int_{0}^{1}g_{1}(x)dx,  \label{sp4}
\end{equation}%
which measures total quark spin contribution to the proton spin. For the $%
g_{1}$ from Eq. (\ref{sp10}) this integral reads%
\begin{equation}
\Gamma _{1}=\frac{1}{2}\int \Delta G(p_{0})\left( \frac{1}{3}+\frac{2m}{%
3p_{0}}\right) d^{3}p.  \label{sp17}
\end{equation}%
Dependence of both \ the integrals (\ref{sp24}) and (\ref{sp17}) on
intrinsic motion is controlled by the same term $\left( 1/3+2m/3p_{0}\right) 
$, which in both the cases has origin in the covariant kinematics of the
particle with $s=1/2$. In fact, the \ procedures for calculation of these
integrals are based on the two different representations of the solutions of
\ Dirac equation: the plane waves (\ref{sp8}) and spherical waves (\ref{sp22}%
). It is apparent that for the scenario of massless quarks $\left( m\ll
p_{0}\right) $, due to necessary presence of the orbital motion, both the
integrals $\Gamma _{1}$ and $\left\langle s\right\rangle $ will be roughly
three times less, than for the scenario of massive and static quarks $\left(
m\simeq p_{0}\right) $. We discussed this effect in the context of
experimental data in \cite{zav05}.

What is the underlying physics behind the interplay between the spin and
orbital momentum? Actually, speaking about the spin of the particle
represented by the state (\ref{sp8}),\ one should take into account:

\textit{a)} Definite projection of the spin in the direction $\mathbf{n}$ is
well-defined quantum number only for the particle at rest ($p=0$) or for the
particle moving in the the direction $\mathbf{n}$, i.e. $\mathbf{p/}p\mathbf{%
=\pm n}$. In these cases we have%
\begin{equation}
s=u^{\dagger }\left( \mathbf{p},\lambda \mathbf{n}\right) \mathbf{n\Sigma }%
u\left( \mathbf{p},\lambda \mathbf{n}\right) =\pm 1/2.  \label{gp32}
\end{equation}

\textit{b)} In other cases, as shown in the Appendix \ref{app3}, only
inequality%
\begin{equation}
\left\langle s\right\rangle =\left\vert u^{\dagger }\left( \mathbf{p}%
,\lambda \mathbf{n}\right) \mathbf{n\Sigma }u\left( \mathbf{p},\lambda 
\mathbf{n}\right) \right\vert <1/2  \label{GP33}
\end{equation}%
is satisfied. Roughly speaking, the result of measuring of the spin (of a
quark) depends on its momentum in the defined reference frame (proton rest
frame). This obvious effect acts also in the states, which are represented
by the superposition of the plane waves (\ref{sp8}) with different momenta $%
\mathbf{p}$ and resulting in $\left\langle \mathbf{p}\right\rangle =0$, but $%
\left\langle \mathbf{p}^{2}\right\rangle >0$. In \cite{zav1} we showed, that
averaging of the spin projection (\ref{GP33}) over the spherical momentum
distribution gives the result equivalent to (\ref{sp17}). The state (\ref%
{sp21}) can be also decomposed into plane waves having spherical momentum
distribution and the spin mean value given by Eq. (\ref{sp24}). Well-defined
quantum numbers $j=j_{z}=1/2$ imply, that the spin reduction due to
increasing intrinsic kinetic energy \ is compensated by the increasing
orbital momentum.

\section{Summary and conclusion}

We studied covariant version of the QPM with spherically symmetric
distributions of \ the quark momentum in the proton rest frame. The main
results obtained in this paper can be summarized as follows.

1) Relations between the structure functions $F_{2},g_{1}$ and corresponding
3D quark momentum distributions $G^{\pm }(p)=G(p)\pm \Delta G(p)$ were
obtained. In this way the momentum distributions of positively and
negatively polarized quarks $G^{\pm }(p)=\sum_{q}e_{q}^{2}G_{q}^{\pm }(p)$
are calculated from the experimentally measured structure functions $F_{2}$
and $g_{1}$. At the same time the partial structure functions $F_{2}$
related to the subsets of quarks, which are described by the distributions $%
G^{\pm }$ and $\Delta G$, were obtained. The momentum distributions are
spherically symmetric, it follows that corresponding longitudinal and
transversal distributions are accessible as well. Results of the calculation
suggest:

\textit{a)} Character of the function $\Delta F_{2}=F_{2}^{+}-F_{2}^{-}$ is
similar to that of the function $F_{2val}$. This suggests, that the dominant
contribution to the proton spin comes from the valence quarks.

\textit{b)} Comparison of $\ $the functions $G^{\pm }$ with $G$ and/or $%
F_{2}^{\pm }$ with $F_{2}$ suggest, that in region of higher intrinsic
energy (and/or higher $x$) the partons with positive polarization strongly
dominate, whereas partons with negative polarization are present rather only
in the lower energy region.

2) We showed, that important role of the quark orbital momentum emerges as a
direct consequence of a covariant description. Since in relativistic case
only the total angular momentum \ $\mathbf{j=l+s}$\ is well-defined quantum
number, there arises some interplay between its spin and orbital parts. For
the quark in the state with definite projection $j_{z}=1/2$ in the proton
rest frame, as a result of this interplay, its spin part is reduced in favor
of the orbital one. The role of orbital motion increases with the rate of
quark intrinsic motion; for $p\gg m$\ its fraction reaches $\left\langle
l_{z}\right\rangle =2/6$ whereas $\left\langle s_{z}\right\rangle =1/6$
only. Simultaneously, this effect is truly reproduced also in the formalism
of\ structure functions.\bigskip 

\textbf{Acknowledgments}

\textit{I would like to thank Anatoli Efremov and Oleg Teryaev for many
useful discussions and valuable comments.}

\appendix

\section{Structure functions in the approach of infinite momentum frame}

\label{app0}The necessary condition for obtaining equalities (\ref{sp11a}) -
(\ref{sp11b}) is the covariant relation%
\begin{equation}
p_{\alpha }=yP_{\alpha },  \label{b2}
\end{equation}%
which implies%
\begin{equation}
m=yM  \label{b3}
\end{equation}%
and $\mathbf{p}=0$ in the proton rest frame and $p_{T}=0$ in the IMF.

For calculation of the integrals (\ref{sp9c}) and (\ref{sp9d}) in the IMF
approach one can substitute $p$ by $yP$ and $d^{3}p/p_{0}$ by $\pi
dp_{T}^{2}dy/y$. Then, after some algebra the structure functions (\ref{sp9b}%
) read%
\begin{equation}
F_{1}(x)=\frac{1}{2}Mx\int G\left( yM\right) \delta \left( y-x\right) \pi
dp_{T}^{2}\frac{dy}{y},\qquad F_{2}(x)=Mx^{2}\int G\left( yM\right) \delta
\left( y-x\right) \pi dp_{T}^{2}\frac{dy}{y}.  \label{b6}
\end{equation}%
Since the approximation (\ref{b2}) implies sharply peaked distribution at $%
p_{T}^{2}\rightarrow 0$, one can identify%
\begin{equation}
MG_{q}\left( yM\right) \pi dp_{T}^{2}=q(y)  \label{b7}
\end{equation}%
and then the Eqs. (\ref{sp11a}) and (\ref{b6}) after integrating are
equivalent.

In the same way the equalities (\ref{cra31})-(\ref{cr31}) can be modified.
Taking into account that $pS\rightarrow yPS=0$, one obtain%
\begin{equation}
g_{1}(x)=\frac{m}{2}\int \Delta G\left( yM\right) \delta \left( y-x\right)
\pi dp_{T}^{2}\frac{dy}{y},\qquad g_{2}(x)=0.  \label{b8}
\end{equation}%
If we put%
\begin{equation}
M\Delta G_{q}\left( yM\right) \pi dp_{T}^{2}=\Delta q(y)  \label{b9}
\end{equation}%
\ and take into account Eq. (\ref{b3}), then it is obvious, that the Eqs. (%
\ref{sp11b}) and (\ref{b8}) are equivalent.

\section{Proof of the relation ({\protect\ref{GP19}})}

\label{app1}In the paper \cite{zav2} we proved relation%
\begin{equation}
\frac{V_{j}^{\prime }(x)}{V_{k}^{\prime }(x)}=\left( \frac{x}{2}+\frac{%
x_{0}^{2}}{2x}\right) ^{j-k};\qquad x_{0}=\frac{m}{M},  \label{a1}
\end{equation}%
which for $m\rightarrow 0$ implies%
\begin{equation}
V_{0}(x)=\frac{1}{2}\left( xV_{-1}(x)+\int_{0}^{x}V_{-1}(y)dy\right) .
\label{a2}
\end{equation}%
After inserting $V_{0}$ from this relation to Eq. (\ref{gp17}) one gets%
\begin{eqnarray}
g_{1}(x) &=&\frac{1}{2}\left( xV_{-1}(x)+\int_{0}^{x}V_{-1}(y)dy\right)
\label{a3} \\
&&-2x^{2}\left( \int_{x}^{1}\frac{V_{-1}(y)}{y^{2}}dy+\int_{x}^{1}\frac{1}{%
y^{3}}\int_{y}^{1}V_{-1}(z)dzdy\right)  \nonumber \\
&&+\frac{1}{2}x\left( \int_{x}^{1}\frac{V_{-1}(y)}{y}dy+\int_{x}^{1}\frac{1}{%
y^{2}}\int_{y}^{1}V_{-1}(z)dzdy\right) .  \nonumber
\end{eqnarray}%
The double integrals can be reduced by integration by parts with the use of
\ formula%
\begin{equation}
\int_{x}^{1}a(y)\left( \int_{y}^{1}b(z)dz\right) dy=\int_{x}^{1}\left(
A(y)-A(x)\right) b(y)dy;\qquad A^{\prime }(x)=a(x),  \label{a4}
\end{equation}%
then the relation (\ref{a3}) is simplified:%
\begin{equation}
g_{1}(x)=\frac{1}{2}xV_{-1}(x)-x^{2}\int_{x}^{1}\frac{V_{-1}(y)}{y^{2}}dy.
\label{a5}
\end{equation}%
In the next step we extract $V_{-1}$ from this relation. After the
substitution $V(x)=V_{-1}(x)/x$ the relation reads%
\begin{equation}
\frac{g_{1}(x)}{x^{2}}=\frac{1}{2}V(x)-\int_{x}^{1}\frac{V(y)}{y}dy,
\label{a6}
\end{equation}%
which implies the differential equation for $V(x)$:%
\begin{equation}
\frac{1}{2}V^{\prime }(x)+\frac{V(x)}{x}=\left( \frac{g_{1}(x)}{x^{2}}%
\right) ^{\prime }.  \label{a7}
\end{equation}%
The corresponding homogeneous equation%
\begin{equation}
\frac{1}{2}V^{\prime }(x)+\frac{V(x)}{x}=0  \label{a8}
\end{equation}%
gives the solution%
\begin{equation}
V(x)=\frac{C}{x^{2}},  \label{a9}
\end{equation}%
which after inserting to Eq. (\ref{a7}) gives%
\begin{equation}
C^{\prime }(x)=2x^{2}\left( \frac{g_{1}(x)}{x^{2}}\right) ^{\prime }.
\label{a10}
\end{equation}%
After integration one easily gets the relation inverse to Eq. (\ref{a5}):%
\begin{equation}
V_{-1}(x)=\frac{2}{x}\left( g_{1}(x)+2\int_{x}^{1}\frac{g_{1}(y)}{y}%
dy\right) ,  \label{a11}
\end{equation}%
which coincides with Eq. (\ref{GP19}).

\section{Proof of the relation ({\protect\ref{GP26}})}

\label{app2}The relation (\ref{gp21}) implies%
\begin{equation}
\int \Delta G(p)d^{3}p=\int_{0}^{1}\left( 3g_{1}(x)+2\int_{x}^{1}\frac{%
g_{1}(y)}{y})dy-xg_{1}^{\prime }(x)\right) dx  \label{a12}
\end{equation}%
and%
\begin{equation}
\int p\Delta G(p)d^{3}p=\frac{M}{2}\int_{0}^{1}\left(
3xg_{1}(x)+2x\int_{x}^{1}\frac{g_{1}(y)}{y})dy-x^{2}g_{1}^{\prime
}(x)\right) dx;\qquad x=\frac{2p}{M}.  \label{a13}
\end{equation}%
If one denotes%
\begin{equation}
\Gamma _{1}=\int_{0}^{1}g_{1}(x)dx,\qquad \Gamma
_{2}=\int_{0}^{1}xg_{1}(x)dx,  \label{a14}
\end{equation}%
then integration by parts gives%
\begin{equation}
\int_{0}^{1}\int_{x}^{1}\frac{g_{1}(y)}{y})dydx=\Gamma _{1},\qquad
\int_{0}^{1}xg_{1}^{\prime }(x)dx=-\Gamma _{1}  \label{a15}
\end{equation}%
and%
\begin{equation}
\int_{0}^{1}2x\int_{x}^{1}\frac{g_{1}(y)}{y})dydx=\Gamma _{2},\qquad
\int_{0}^{1}x^{2}g_{1}^{\prime }(x)dx=-2\Gamma _{2}.  \label{a16}
\end{equation}%
Now, one can easily express the ratio%
\begin{equation}
\frac{\int p\Delta G(p)d^{3}p}{\int \Delta G(p)d^{3}p}=\frac{M}{2}\frac{%
\Gamma _{2}}{\Gamma _{1}},  \label{a17}
\end{equation}%
in this way the relation ({\ref{GP26}}) is proved.

\section{Proof of the relation ({\protect\ref{GP33}})}

\label{app3}With the use of rule%
\begin{equation}
\mathbf{p\sigma \cdot n\sigma }+\mathbf{n\sigma \cdot p\sigma =2pn}
\label{a18}
\end{equation}%
the term in Eq. ({\ref{GP33}}) can be modified as

\begin{eqnarray}
u^{\dagger }\left( \mathbf{p},\lambda \mathbf{n}\right) \mathbf{n\Sigma }%
u\left( \mathbf{p},\lambda \mathbf{n}\right) &=&\frac{1}{2N}\phi _{\lambda 
\mathbf{n}}^{\dagger }\left( \mathbf{n\sigma +}\frac{\mathbf{p}\mathbf{%
\sigma \cdot n\sigma \cdot p}\mathbf{\sigma }}{\left( p_{0}+m\right) ^{2}}%
\right) \phi _{\lambda \mathbf{n}}  \label{a19} \\
&=&\frac{1}{2N}\phi _{\lambda \mathbf{n}}^{\dagger }\left( \mathbf{n\sigma +}%
\frac{\mathbf{p}\mathbf{\sigma \cdot }\left( -\mathbf{p\sigma \cdot n\sigma
+2pn}\right) }{\left( p_{0}+m\right) ^{2}}\right) \phi _{\lambda \mathbf{n}}
\nonumber \\
&=&\frac{1}{2N}\phi _{\lambda \mathbf{n}}^{\dagger }\left( \mathbf{n\sigma }%
\left( \mathbf{1-}\frac{\mathbf{p}^{2}}{\left( p_{0}+m\right) ^{2}}\right) 
\mathbf{+}\frac{\mathbf{2pn\cdot p}\mathbf{\sigma }}{\left( p_{0}+m\right)
^{2}}\right) \phi _{\lambda \mathbf{n}}  \nonumber \\
&=&\frac{1}{2p_{0}}\phi _{\lambda \mathbf{n}}^{\dagger }\left( m\cdot 
\mathbf{n\sigma +}\frac{\mathbf{pn\cdot p}\mathbf{\sigma }}{p_{0}+m}\right)
\phi _{\lambda \mathbf{n}}.  \nonumber
\end{eqnarray}%
Since%
\begin{equation}
\left| \phi _{\lambda \mathbf{n}}^{\dagger }\mathbf{n\sigma }\phi _{\lambda 
\mathbf{n}}\right| =1,\qquad \left| \phi _{\lambda \mathbf{n}}^{\dagger }%
\mathbf{p\sigma }\phi _{\lambda \mathbf{n}}\right| \leq p,\qquad \mathbf{pn=}%
p\mathbf{\cos \alpha ,}  \label{a20}
\end{equation}%
it follows%
\begin{equation}
\left| u^{\dagger }\left( \mathbf{p},\lambda \mathbf{n}\right) \mathbf{%
n\Sigma }u\left( \mathbf{p},\lambda \mathbf{n}\right) \right| \leq \frac{1}{%
2p_{0}}\left( m\mathbf{+}\frac{p^{2}}{p_{0}+m}\right) =\frac{1}{2}.
\label{a21}
\end{equation}%
Obviously 
\begin{equation}
\left| u^{\dagger }\left( \mathbf{p},\lambda \mathbf{n}\right) \mathbf{%
n\Sigma }u\left( \mathbf{p},\lambda \mathbf{n}\right) \right| =\frac{1}{2}
\label{a22}
\end{equation}%
only for $\mathbf{p/}p\mathbf{=\pm n}$ or $p=0\mathbf{.}$


\begin{thebibliography}{99}
\bibitem{e142} E142 Collaboration, P.L. Anthony \textit{et. al.}, Phys. Rev.
D \textbf{54}, 6620 (1996).

\bibitem{e154} E154 Collaboration, K. Abe \textit{et. al}., Phys. Rev. Lett. 
\textbf{79}, 26 (1997).

\bibitem{her1} HERMES Collaboration, K. Ackerstaff \textit{et. al}., Phys.
Lett. B\textbf{\ 404}, 383 (1997).

\bibitem{her2} HERMES Collaboration, A. Airapetian \textit{et. al}., Phys.
Lett. B\textbf{\ 442}, 484 (1998).

\bibitem{adeva} Spin Muon Collaboration, B. Adeva \textit{et. al}., Phys.
Rev. D\textbf{\ 58}, 112001 (1998).

\bibitem{e143} E143 Collaboration, K. Abe \textit{et. al}., Phys. Rev. D 
\textbf{58}, 112003 (1998).

\bibitem{E155} E155 Collaboration, P. Anthony \textit{et al.}, Phys. Lett. B 
\textbf{493}, 19 (2000); Phys. Lett. B \textbf{553}, 18 (2003).

\bibitem{spin04} \textit{Proceedings \ of the 16th International Spin
Physics Symposium - SPIN 2004}, Trieste, Italy, 10-16 October 2004, edited
by Franco Bradamante \textit{et. al.} (World Scientific Publishing Co.Pte.
Ltd.).

\bibitem{efrem} M.~Anselmino, A.~Efremov, E.~Leader, Phys. Rep. \textbf{261}%
, 1 (1995).

\bibitem{zav0} P.~Zavada, Phys. Rev. D\textbf{\ 55}, 4290 (1997).

\bibitem{zav1} P.~Zavada, Phys. Rev. D\textbf{\ 65}, 0054040 (2002).

\bibitem{zav2} P.~Zavada, Phys. Rev. D\textbf{\ 67}, 014019 (2003).

\bibitem{zav05} P.\ Zavada, arXiv: hep-ph/0511142.

\bibitem{tra} A.V. Efremov, O.V. Teryaev and P. Zavada, Phys. Rev. D\textbf{%
\ 70}, 054018 (2004), \ arXiv: hep-ph/0512034.

\bibitem{sehgal} L.M. Sehgal, Phys. Rev. D\textbf{\ 10}, 1663 (1974); D%
\textbf{\ 10}, 2016 (E) (1974).

\bibitem{ratc} P.G. Ratcliffe, Phys. Lett. B \textbf{192}, 180 (1987).

\bibitem{abbas} A. Abbas, J.Phys. G, Nucl. Part. Phys. \textbf{16}, L21-L26
(1990); \textbf{15}, L73-L77 (1989).

\bibitem{kep} J. Keppler and H.M. Hofmann, Phys. Rev. D\textbf{\ 51}, 3936
(1995).

\bibitem{casu} M. Casu and L.M. Sehgal, Phys. Rev. D\textbf{\ 55}, 2644
(1996).

\bibitem{bqma} B.-Q. Ma and I. Schmidt, Phys. Rev. D\textbf{\ 58}, 096008
(1998).

\bibitem{brod} S. J. Brodsky, Dae Sung Hwang, B.-Q. Ma and I. Schmidt,
Nucl.Phys.B \textbf{593, }311 (2001).

\bibitem{waka} M. Wakamatsu and T. Watabe, Phys. Rev. D\textbf{\ 62}, 054009
(2000).

\bibitem{waka1} M. Wakamatsu and Y. Nakakoji, Phys. Rev. D \textbf{74}
054006 (2006).

\bibitem{song} X. Song, Int. J. Mod. Phys. A \textbf{16}, 3673 (2001).

\bibitem{ji} X. Ji, J.-P. Ma and F. Yuan, Nucl. Phys. B \textbf{652}, 383
(2003).

\bibitem{liang} Z. Liang and T. Meng, Z. Phys. A \textbf{344}, 171 (1992).

\bibitem{lali} L.D. Landau, E.M. Lifshitz \textit{et al.}, Quantum
Electrodynamics (Course of Theoretical Physics, vol. 4), Elsevier Science
Ltd., 1982.
\end{thebibliography}
\end{document}